\documentclass[11pt]{article}
\usepackage{amsmath,amssymb,mathtools}
\usepackage[margin=1in]{geometry}
\usepackage{graphicx}
\usepackage{color}
\usepackage{tikz}
\usepackage{subcaption}
\usetikzlibrary{arrows.meta}

\def\bmsigma{\mbox{\boldmath $\sigma$}}
\def\bmvarepsilon{\mbox{$\boldsymbol \varepsilon$}}

\def\veps{\mbox{$\varepsilon$}}

\def\be{\begin{equation}}
\def\ee{\end{equation}}
\def\ba{\begin{array}}
\def\ea{\end{array}}

\newcommand{\epsg}{\varepsilon^{g}}

\title{\bf When Volumetric Growth Selects Surface Growth}

\author{
Rohan Abeyaratne\thanks{Department of Mechanical Engineering,
Massachusetts Institute of Technology, Cambridge, MA 02139, USA.
\texttt{rohan@mit.edu}}
\and
Roberto Paroni\thanks{Department of Civil and Industrial Engineering,
University of Pisa, Largo Lucio Lazzarino 1, 56122 Pisa, Italy.
\texttt{roberto.paroni@unipi.it}}
\and
Marco Picchi Scardaoni\thanks{Department of Civil and Industrial Engineering,
University of Pisa, Largo Lucio Lazzarino 1, 56122 Pisa, Italy.
\texttt{marco.picchiscardaoni@ing.unipi.it}}
}

\date{\today}

\begin{document}

\maketitle

\begin{abstract}
We investigate the relationship between volumetric and surface growth within a recently proposed optimization-driven framework for linearly elastic solids. In this approach, growth is not prescribed through an evolution law; instead, the growth distribution is determined as the solution of a constrained optimization problem.

Focusing on processes driven by the minimization of the work performed by external loads in one-dimensional and axisymmetric settings, we derive explicit analytical solutions for the resulting growth distributions. Although growth is initially formulated as a volumetric process through a distributed growth strain tensor, we show that the optimal growth distributions are singular and concentrate on boundaries or internal interfaces.

These results provide a variational mechanism through which, under certain conditions, surface growth is selected as the optimal realization of volumetric growth.
\end{abstract}

\section{Introduction}

Growth in solids may occur either through the addition of material throughout the bulk or through accretion at the boundary. These two mechanisms, commonly referred to as volumetric growth and surface growth, respectively, play a central role in a wide variety of biological and physical systems, including morphogenesis, tumor development, bone remodeling, additive manufacturing, crystal growth, and the evolution of living tissues. Although both processes involve the generation of new material and the consequent modification of geometry and stress, they have traditionally been described within distinct theoretical frameworks.

Volumetric growth is commonly modeled through the introduction of a growth tensor whose evolution is prescribed by an evolution law. Following the seminal work of Skalak et al.~\cite{Skalak1982} and the influential morphoelastic framework of Rodriguez et al.~\cite{Rodriguez1994}, growth is usually represented through a multiplicative decomposition of the deformation gradient into elastic and growth components. Subsequent developments incorporated stress- and strain-driven growth laws, biochemical stimuli, residual stresses, geometric incompatibility, and growth-induced instabilities; representative contributions include \cite{DiCarloQuiligotti,AmbrosiMollica,ChenHoger,GorielyBenAmar}, while broader accounts can be found in \cite{Goriely,Taber}.

Surface growth, by contrast, is typically formulated as a boundary accretion process in which new material is deposited on the evolving surface of the body. Such approaches commonly rely on moving-boundary descriptions and kinetic relations governing the accretion velocity. Representative contributions include the works of Tomassetti et al.~\cite{Tomassetti}, Zurlo and Truskinovsky~\cite{ZurloTruskinovsky1,ZurloTruskinovsky2,ZurloTruskinovsky3}, Naghibzadeh et al.~\cite{Naghibzadeh}, and Renzi et al.~\cite{Renzi}.

While volumetric and surface growth are often modeled separately, there are indications that the two mechanisms may be more closely related than their traditional descriptions suggest. In particular, Di Carlo~\cite{DiCarlo2005} argued that surface growth may be interpreted as the singular limit of an increasingly localized volumetric growth process, suggesting that both mechanisms can be regarded as manifestations of a common underlying phenomenon.

Motivated by the observation that many biological systems appear to evolve toward configurations that optimize specific functional or energetic objectives while satisfying physical and physiological constraints, in \cite{Abeyaratne1,Abeyaratne2} we proposed a framework in which growth is not prescribed through an evolution law but instead emerges as the solution of a constrained optimization problem. A variational formulation based on these ideas was first introduced for surface growth \cite{Abeyaratne1}, where accretion was selected by minimizing the structural compliance of a growing beam. More recently, the same philosophy was extended to volumetric growth in linearized elasticity \cite{Abeyaratne2}.

The purpose of the present work is to investigate {\it the relationship between} volumetric and surface growth within this optimization-driven framework. We focus on problems in which growth is determined by minimizing the work performed by externally applied loads under prescribed mass and equilibrium constraints. Here, the  added material is allocated so as to induce a displacement that most effectively opposes the loading, producing the most negative  work. One-dimensional and axisymmetric settings are particularly attractive because they permit explicit analytical calculations while retaining some of the essential features of higher-dimensional growth processes. While the present analysis is carried out within the linearly elastic framework, ongoing work suggests that similar results carry over to the nonlinear theory. 

The central result of the paper is that, although growth is initially formulated as a volumetric process through a distributed growth strain tensor, the optimal solutions are singular. More precisely, the minimizing growth distributions concentrate on lower-dimensional subsets of the body and converge, in the sense of measures, to growth strains supported on boundaries or lower-dimensional internal sets. In the case of a uniformly loaded annulus, the optimal growth localizes at the inner and outer boundaries, whereas for inward radial loading it concentrates at the inner boundary alone.  It also selects the component of the growth strain, i.e. radial versus hoop. Thus, the optimization problem spontaneously selects growth mechanisms that are effectively of surface type.

These results provide a concrete realization of the idea that surface growth may emerge as a limiting form of volumetric growth in certain settings. Unlike classical approaches, where the distinction between bulk and surface growth is imposed a priori at the modeling stage, here it arises naturally from the optimization process itself. The present work therefore provides a variational mechanism through which surface accretion can be interpreted as the optimal realization of volumetric growth under suitable mechanical objectives.

The paper is organized as follows. Section~2 briefly recalls the optimization-driven formulation of volumetric growth proposed in \cite{Abeyaratne2}. Section~3 studies growth in a one-dimensional bar problem, which serves as a prototype illustrating growth localization. Section~4 analyzes axisymmetric plane-strain growth in an annular body and derives explicit expressions for the work functional. The resulting optimal growth distributions are then characterized for different loading conditions, revealing the emergence of measure-valued solutions concentrated on lower-dimensional sets. Finally, Section~5 contains some concluding remarks.

\section{Volumetric growth driven by an optimality criterion}

In this section we briefly recall the optimization-driven formulation of volumetric growth introduced in \cite{Abeyaratne2}. Rather than reproducing the general theory in full, we focus on the ingredients that are relevant to the problems studied in the following sections.

Consider a linearly elastic body occupying a reference configuration $\Omega$. Within the framework of linearized elasticity, growth is described through a symmetric tensor field $\bmvarepsilon^g$, referred to as the growth strain. The total strain is decomposed additively as
\[
\bmvarepsilon=\frac 12 (\nabla \mathbf{u}+\nabla\mathbf{u}^{\mathsf T})=\bmvarepsilon^e+\bmvarepsilon^g,
\]
where $\mathbf{u}$ and $\bmvarepsilon^e$ are the displacement and the elastic part of the strain, respectively. The stress tensor is
\[
\bmsigma=\mathbb C(\bmvarepsilon-\bmvarepsilon^g),
\]
where $\mathbb C$ denotes the elasticity tensor.

The quantity $\operatorname{tr}\bmvarepsilon^g$ represents the local volumetric change associated with growth. Accordingly, the total amount of material added to the body is measured by
\[
\int_\Omega \operatorname{tr}\bmvarepsilon^g\,dx .
\]
Rather than prescribing the evolution of $\bmvarepsilon^g$ through a evolution law, we assume that growth is selected by an optimality principle. More precisely, among all admissible growth distributions satisfying a prescribed mass constraint and consistent with the mechanical equations, the body chooses the one that minimizes a suitable objective functional.

In the present paper the objective functional is the work performed by the applied loads. With this choice, the body allocates a prescribed amount of added material so that the induced displacement most effectively opposes the loading, producing the most negative work.
Denoting by $\mathbf{u}$ the displacement field and by $W(\mathbf{u},\bmvarepsilon^g)$ the work of the external loads, the growth problem takes the abstract form
\[
\inf_{u,\varepsilon^g} W(\mathbf{u},\bmvarepsilon^g),
\]
subject to

\begin{itemize}
\item the equations of mechanical equilibrium,
\item the constitutive relation
\[
\bmsigma=\mathbb C(\bmvarepsilon-\bmvarepsilon^g),
\]
\item a prescribed amount of added material
\[
\int_\Omega \operatorname{tr}\bmvarepsilon^g\,dx=\Gamma;
\]
assuming unit mass density for the added material, $\Gamma$ represents the total mass added to the body,
\item and the irreversibility condition
\[
\bmvarepsilon^g\mathbf{v}\cdot\mathbf{v}\ge 0\qquad\forall\mathbf{v},
\]
which expresses the fact that growth can only add material in the ``intermediate'' configuration.
\end{itemize}

The central feature of this formulation is that the growth strain is not assigned a priori nor by a phenomenological kinetic law. Instead, it emerges as the solution of a constrained optimization problem. In the general framework developed in \cite{Abeyaratne2}, growth evolves incrementally in time and is determined at each step by solving a constrained minimization problem. Here, for simplicity, we restrict attention to the corresponding static optimization problem, which is sufficient to reveal the localization mechanisms of interest.

\section{Volumetric growth of a bar}
Consider a linearly elastic bar occupying the interval $[0,\ell]$, fixed at both ends and subjected to a distributed axial load $p(x)$ per unit length.

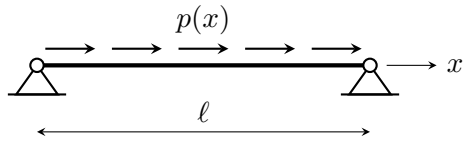
\begin{center}
\begin{tikzpicture}[scale=1.1,>=stealth]
  \def\L{4}

  \draw[ultra thick] (0,0) -- (\L,0);

   \draw[thick] (-0.25,-0.35) -- (0.25,-0.35) -- (0,0) -- cycle;
  \draw[thick] (-0.35,-0.35) -- (0.35,-0.35);
 \draw[thick, fill=white] (0,0) circle (0.08);

  \draw[thick] (\L-0.25,-0.35) -- (\L+0.25,-0.35) -- (\L,0) -- cycle;
  \draw[thick] (\L-0.35,-0.35) -- (\L+0.35,-0.35);
 \draw[thick, fill=white] (\L,0) circle (0.08);
  \draw[->] (\L+0.2,0) -- (\L+0.8,0) node[right] {$x$};

  \foreach \x in {0.1,0.9,1.7,2.5,3.3}
    \draw[->,thick] (\x,0.2) -- (\x+0.6,0.2);
  \node[above] at (2,0.25) {$p(x)$};

  \draw[<->] (0,-0.8) -- (\L,-0.8);
  \node[above] at (2,-0.8) {$\ell$};
\end{tikzpicture}

%
%
%
%
%
\end{center}
Letting $u$ denote the axial displacement we have that
\begin{equation*}\label{bdrycond}
u(0)=0\quad\mbox{and}\quad u(\ell)=0.
\end{equation*}
Denoting by $\varepsilon$ the total strain, by $\varepsilon^g$ the  growth strain, by $N$ the axial force,  and by $EA$ the axial stiffness, we have
\begin{equation*}\label{laws0}
\varepsilon(x)=u'(x),\qquad\quad \varepsilon^g(x)\ge 0, \qquad\quad N(x)=EA \big(\varepsilon(x)-\varepsilon^g(x)\big)\qquad \text{for} \quad 0 \leq x \leq \ell.
\end{equation*}
Also, equilibrium dictates that 
$$
N'(x)+p(x)=0\qquad \text{for} \quad 0 \leq x \leq \ell.
$$ 
We assume that the growth of the bar, i.e., the choice of the function $\varepsilon^g$, is
governed by the following optimality criterion: the beam selects $\varepsilon^g$ so as to minimize the work done by the loads
$$
W(u,\varepsilon^g):=\int_0^\ell p(x) u(x)\,dx,
$$
subjected to the constraint
$$
\int_0^\ell \varepsilon^g(x)\,dx=\Gamma,
$$
where $\Gamma$ is a  positive constant.  The growth problem therefore is:
\begin{equation}\label{pbm1d}
\begin{cases}
\displaystyle \inf_{u, \varepsilon^g}  & \displaystyle W(u,\varepsilon^g)=\int_0^\ell p(x) u(x)\,dx,\\
&\mbox{subject to}\\
& N'+p=0,\\
& N=EA (\varepsilon-\varepsilon^g),\\
& \varepsilon=u',\\
& \varepsilon^g\ge 0,\\
& \displaystyle \int_0^\ell \varepsilon^g(x)\,dx=\Gamma,\\
& u(0)=0\mbox{ and } u(\ell)=0.
\end{cases}
\end{equation}
\subsection{Determination of the displacement and axial force fields}
By using \eqref{pbm1d}$_2$, we have
$$
N(x)=N_0-\int_0^x p(s)\,ds \qquad \text{for} \quad 0 \leq x \leq \ell,
$$
where $N_0=N(0)$ is a constant.
Denoting by $\varepsilon^e=\varepsilon-\varepsilon^g$ the elastic strain, we deduce from \eqref{pbm1d}$_3$
$$
    \varepsilon^e(x) = \frac{N(x)}{EA}=\frac{N_0}{EA}-\frac{1}{EA}\int_0^x p(s)\,ds,
$$
and so the total strain can be written as
\begin{equation*}
\label{eq:1}
    \varepsilon(x) = u'(x)=\varepsilon^e(x)+\epsg(x)=\frac{N_0}{EA}-\frac{1}{EA}\int_0^x p(s)\,ds+\epsg(x) \qquad \text{for} \quad 0 \le x \le \ell.
\end{equation*}
Integrating and using $u(0)=0$ gives
$$
    u(x)= \frac{N_0 x}{EA}-\frac{1}{EA}\int_0^x\int_0^\xi  p(s)\,ds\, d\xi+\int_0^x \epsg(s)\,ds \qquad \text{for} \quad 0 \leq x \leq \ell.
$$
The boundary condition $u(\ell)=0$ yields
\begin{equation*}
    0=\frac{N_0 \ell}{EA}-\frac{1}{EA}\int_0^\ell\int_0^\xi  p(s)\,ds\, d\xi+\int_0^\ell \epsg(s)\,ds  \quad \implies \quad
        N_0=\frac{1}{\ell}\int_0^\ell\int_0^\xi  p(s)\,ds\, d\xi-\frac{EA\Gamma}{\ell},
    \label{eq:2}
\end{equation*}
where we used \eqref{pbm1d}$_5$.
Thus the internal normal force field is
\begin{equation}
    N(x)=\frac{1}{\ell}\int_0^\ell\int_0^\xi  p(s)\,ds\, d\xi-\frac{EA\Gamma}{\ell} -\int_0^x p(s)\,ds \qquad \text{for} \quad 0 \leq x \leq \ell, 
    \label{N}
\end{equation}
and the displacement is given by
\begin{equation}
    u(x)= \Big(\frac{1}{EA\ell}\int_0^\ell\int_0^\xi  p(s)\,ds\, d\xi-\frac{\Gamma}{\ell}\Big)x-\frac{1}{EA}\int_0^x\int_0^\xi  p(s)\,ds\, d\xi+\int_0^x \epsg(s)\,ds \qquad \text{for} \quad 0 \leq x \leq \ell.    
    \label{u1d}
\end{equation}
We remark that the axial force $N$ is independent of $\epsg$.
%
%

\subsection{Minimization of the work done by the loads}\label{meassol}
By using the equilibrium equation $N'+p=0$, the work done by the loads writes as
$$
\begin{aligned}
W(u,\varepsilon^g)&=\int_0^\ell p u\,dx=-\int_0^\ell N' u\,dx=-\int_0^\ell [(N u)'- Nu']\,dx=\int_0^\ell N\varepsilon\,dx\\
&=\int_0^\ell N(\varepsilon^e+\varepsilon^g)\,dx =\int_0^\ell \big(\frac{N^2}{EA}+N \varepsilon^g\big)\,dx
\end{aligned}
$$
and, by using \eqref{N} and \eqref{pbm1d}$_5$,  we deduce that
$$
\begin{aligned}
 \inf_{u, \varepsilon^g} W(u,\varepsilon^g)&=\int_0^\ell \frac{N^2}{EA}\,dx+ \inf_{\varepsilon^g}\int_0^\ell N \varepsilon^g\,dx\\
 &\ge\int_0^\ell \frac{N^2}{EA}\,dx+ \min_{x\in[0,\ell]}N(x) \inf_{\varepsilon^g} \int_0^\ell  \varepsilon^g\,dx.
\end{aligned}
$$
Hence, by \eqref{pbm1d}$_6$, we find
\begin{equation}
 \inf_{u, \varepsilon^g} W(u,\varepsilon^g)\ge\int_0^\ell \frac{N^2}{EA}\,dx+ \Gamma \min_{x\in[0,\ell]}N(x).
    \label{lowerb1d}
\end{equation}

Equation \eqref{lowerb1d} establishes a lower bound for the work done by the loads. We now examine whether this bound is attainable and characterize the strain $\varepsilon^g$ for which equality is achieved.

Let $x_\circ \in [0,\ell]$ denote a point at which $N$ attains its minimum:
$$
N(x_\circ)=\min_{x\in[0,\ell]}N(x).
$$
For each positive integer $k$, define
\[
B_k=
\begin{cases}
[0,\frac{1}{k}], & \text{if } x_\circ=0,\\[1mm]
\left[x_\circ-\frac{1}{2k},\,x_\circ+\frac{1}{2k}\right], & \text{if } 0<x_\circ<\ell,\\[1mm]
\left[\ell-\frac{1}{k},\,\ell\right], & \text{if } x_\circ=\ell,
\end{cases}
\]
and let
\[
\varepsilon_k^g(x):=k\Gamma\,\chi_{B_k}(x),
\]
where $\chi_{B_k}$ denotes the characteristic function of $B_k$, {\it i.e.,} $\chi_{B_k}(x)=1$ if $x\in B_k$ and $0$ otherwise.
 Clearly,
$\varepsilon_k^g(x)\geq 0$ for all $x\in[0,\ell]$. The function $\varepsilon_k^g$ represents a growth strain uniformly distributed over the interval $B_k$ and vanishing elsewhere. Its magnitude is chosen to be $k\Gamma$, so that the total amount of growth contained in $B_k$ is equal to $\Gamma$:
for every sufficiently large $k$
\[
\int_0^\ell \varepsilon_k^g(x)\,dx=\Gamma.
\]
Thus, $\varepsilon_k^g$ describes a growth process localized within the region $B_k$, with larger values of $k$ corresponding to growth that is distributed over a smaller portion of the bar while preserving the same total growth.

The work done by the loads corresponding to $\varepsilon_k^g$ is
\[
W(u,\varepsilon_k^g)
=
\int_0^\ell \frac{N^2}{EA}\,dx
+\int_0^\ell N\,\varepsilon_k^g\,dx
=
\int_0^\ell \frac{N^2}{EA}\,dx
+\Gamma \frac{1}{1/k}\int_{B_k}N(x)\,dx.
\]
Since the average value of $N$ over $B_k$ converges to $N(x_\circ)$ as $k\to\infty$, it follows that
\[
\lim_{k\to\infty}W(u,\varepsilon_k^g)
=
\int_0^\ell \frac{N^2}{EA}\,dx
+\Gamma N(x_\circ).
\]
Recalling that $N(x_\circ)=\min_{x\in[0,\ell]}N(x)$, we conclude that
\[
\lim_{k\to\infty}W(u,\varepsilon_k^g)
=
\int_0^\ell \frac{N^2}{EA}\,dx
+\Gamma \min_{x\in[0,\ell]}N(x),
\]
showing that the lower bound in \eqref{lowerb1d} is attained in the limit.

To identify the limit of $\varepsilon_k^g$, let $\psi$ be a smooth function on $[0,\ell]$. Then
\[
\lim_{k\to\infty}\int_0^\ell \psi(x)\,\varepsilon_k^g(x)\,dx
=
\Gamma \lim_{k\to\infty} k\int_{B_k}\psi(x)\,dx
=
\Gamma \psi(x_\circ)
=
\Gamma \int_{[0,\ell]}\psi\,d\delta_{x_\circ},
\]
where $\delta_{x_\circ}$ denotes the Dirac measure concentrated at $x_\circ$. Hence,  the ``limit'' of $\varepsilon^g_k$, as $k$ goes to infinity, is the measure $\Gamma \delta_{x_\circ}$.
That is, the sequence $\{\varepsilon_k^g\}$ represents a growth distribution that becomes increasingly concentrated at $x_\circ$, converging (in the sense of measures) to the Dirac mass $\Gamma\delta_{x_\circ}$.

If there are more minimum points of the axial force, the mass $\Gamma$ can be arbitrarily distributed among those points, see Example 3.

\subsection*{Example 1: constant load}
As a first example, we consider $p(x)=p$, with $p$ a fixed positive constant.
By means of \eqref{N} we deduce that
$$
N(x)=p(\frac \ell 2 -x)-\frac{EA\Gamma}{\ell},
$$
and hence
$$N(\ell)=\min_{x\in[0,\ell]} N(x)=-\frac{p\ell} 2 -\frac{EA\Gamma}{\ell}.
$$
Thus,
\[
x_\circ=\ell,
\]
which implies that all the added material is concentrated at the right end of the bar:

\hfill
  \begin{minipage}{0.5\textwidth}
\begin{tikzpicture}[scale=1.1,>=stealth]
  \def\L{4}

  \draw[ultra thick] (0,0) -- (\L,0);

   \draw[thick] (-0.25,-0.35) -- (0.25,-0.35) -- (0,0) -- cycle;
  \draw[thick] (-0.35,-0.35) -- (0.35,-0.35);
 \draw[thick, fill=white] (0,0) circle (0.08);

  \draw[thick] (\L-0.25,-0.35) -- (\L+0.25,-0.35) -- (\L,0) -- cycle;
  \draw[thick] (\L-0.35,-0.35) -- (\L+0.35,-0.35);
 \draw[thick, fill=white] (\L,0) circle (0.08);
\draw[line width=5pt,red] (\L-0.18,0) -- (\L-0.08,0);
  \foreach \x in {0.1,0.9,1.7,2.5,3.3}
    \draw[->,thick] (\x,0.2) -- (\x+0.6,0.2);
  \node[above] at (2,0.25) {$p$};
\end{tikzpicture}
 \end{minipage}
 \quad
   \begin{minipage}{0.3\textwidth}
   $\epsg=\Gamma\delta_\ell.$
 \end{minipage}

This result is physically intuitive: growth localized at the right end pushes the entire bar toward the left opposing the external loads, thereby causing the external loads to perform the largest negative work. Indeed, from \eqref{u1d} we obtain
\[
u(x)=\frac{p}{2EA}x(\ell-x)+\int_0^x \varepsilon^g(s)\,ds-\frac{\Gamma}{\ell}x,
\qquad 0 \le x \le \ell,
\]
and since the integral term vanishes for all $x<\ell$ when the growth strain is concentrated at $x_\circ=\ell$, the displacement is minimized at every interior point of the bar, yielding
\[
u(x)=\frac{p}{2EA}x(\ell-x)-\frac{\Gamma}{\ell}x,
\qquad 0 \le x < \ell.
\]
While growth is described in Example \ref{pbm1d} as a volumetric phenomenon through the distributed field $\varepsilon^g$, the optimal solution corresponds to a singular concentration of growth at the boundary. The resulting measure-valued growth strain is supported on the point $x=\ell$, so that the volumetric growth process effectively  may be considered of surface type.

\subsection*{Example 2:  load positive in  $\boldsymbol{(0,\ell/2)}$ and negative in  $\boldsymbol{(\ell/2,\ell)}$}
Let $p>0$ and
$$
p(x)=\begin{cases}
+p & \mbox{if }0<x<\ell/2\\
-p & \mbox{if }\ell/2<x<\ell.
\end{cases}
$$
Since $N'=-p$, the axial force decreases in $(0,\ell/2)$ and increases on the other half. Hence, the minimum occurs at 
$
x_\circ=\ell/2,
$
which implies that all the added material is concentrated in the middle of the bar:

\hfill
  \begin{minipage}{0.5\textwidth}
\begin{tikzpicture}[scale=1.1,>=stealth]
  \def\L{4}

  \draw[ultra thick] (0,0) -- (\L,0);

   \draw[thick] (-0.25,-0.35) -- (0.25,-0.35) -- (0,0) -- cycle;
  \draw[thick] (-0.35,-0.35) -- (0.35,-0.35);
 \draw[thick, fill=white] (0,0) circle (0.08);

  \draw[thick] (\L-0.25,-0.35) -- (\L+0.25,-0.35) -- (\L,0) -- cycle;
  \draw[thick] (\L-0.35,-0.35) -- (\L+0.35,-0.35);
 \draw[thick, fill=white] (\L,0) circle (0.08);
\draw[line width=5pt,red] (\L/2-0.05,0) -- (\L/2+0.05,0);
  \foreach \x in {0.1,0.7,1.3}
    \draw[->,thick] (\x,0.2) -- (\x+0.5,0.2);
  \node[above] at (1,0.25) {$+p$};
    \foreach \x in {2.1,2.7,3.3}
    \draw[<-,thick] (\x,0.2) -- (\x+0.5,0.2);
  \node[above] at (3,0.25) {$-p$};

\end{tikzpicture}
\end{minipage}
\quad 
  \begin{minipage}{0.3\textwidth}
  $\epsg=\Gamma\delta_{\ell/2}.$
\end{minipage}

Again, the result admits a clear physical interpretation. Growth concentrated at the midpoint of the bar generates an internal expansion that drives the two halves of the bar toward the hinges. As a result, the induced deformation is converted as efficiently as possible into displacements opposing the external loads. Consequently, the loads perform the maximum negative work.

Unlike the previous example, where the optimal growth concentrated at the boundary and could be viewed as a surface mechanism, the optimal growth here is localized at an interior point and acts as an internal source of expansion.

\subsection*{Example 3:  load negative in  $\boldsymbol{(0,\ell/2)}$ and positive in  $\boldsymbol{(\ell/2,\ell)}$}
Let $p>0$ and
$$
p(x)=\begin{cases}
-p & \mbox{if }0<x<\ell/2\\
+p & \mbox{if }\ell/2<x<\ell.
\end{cases}
$$
Arguing as in Example 2 we find that the axial force  takes the minimum values at the two ends, therefore
$
x_\circ\in \{0,\ell\},
$
which implies that all the added material concentrates at either or both ends of the bar:

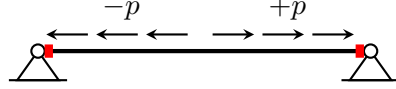
\hfill
  \begin{minipage}{0.4\textwidth}
\begin{tikzpicture}[scale=1.1,>=stealth]
  \def\L{4}

  \draw[ultra thick] (0,0) -- (\L,0);

   \draw[thick] (-0.25,-0.35) -- (0.25,-0.35) -- (0,0) -- cycle;
  \draw[thick] (-0.35,-0.35) -- (0.35,-0.35);
 \draw[thick, fill=white] (0,0) circle (0.08);

  \draw[thick] (\L-0.25,-0.35) -- (\L+0.25,-0.35) -- (\L,0) -- cycle;
  \draw[thick] (\L-0.35,-0.35) -- (\L+0.35,-0.35);
 \draw[thick, fill=white] (\L,0) circle (0.08);
\draw[line width=5pt,red] (\L-0.18,0) -- (\L-0.08,0);
\draw[line width=5pt,red] (0.08,0) -- (0.18,0);
  \foreach \x in {0.1,0.7,1.3}
    \draw[<-,thick] (\x,0.2) -- (\x+0.5,0.2);
  \node[above] at (1,0.25) {$-p$};
    \foreach \x in {2.1,2.7,3.3}
    \draw[->,thick] (\x,0.2) -- (\x+0.5,0.2);
  \node[above] at (3,0.25) {$+p$};

\end{tikzpicture}
\end{minipage}
\quad 
  \begin{minipage}{0.4\textwidth}
  $\epsg=\eta\Gamma\delta_{0}+(1-\eta)\Gamma\delta_{\ell}\quad \forall \eta\in [0,1].$
\end{minipage}

\section{Axisymmetric Plane-Strain Growth}

We consider a body occupying, in its reference configuration, a annular region with inner radius $a$ and outer radius $b$:
\[
\Omega = \{(r,\theta)\,:\, a<r<b\},
\]
where $(r,\theta)$ denote the polar coordinates.
We assume a radially symmetric displacement field of the form
\[
\mathbf{u}(r)=u(r)\,\mathbf{e}_r,
\]
where $\mathbf{e}_r$ is the radial unit vector. The corresponding infinitesimal strain tensor has components
\[
\varepsilon_{rr}=u', \qquad \varepsilon_{\theta\theta}=\frac{u}{r}, \qquad \varepsilon_{r\theta} = 0, 
\]
where \( (\cdot)' = d(\cdot)/dr \).
The growth strain is assumed to be axisymmetric and is represented by
\[
\varepsilon^g_{rr}(r),
\qquad
\varepsilon^g_{\theta\theta}(r),
\qquad
\varepsilon^g_{r\theta}(r)=0.
\]
Assuming that the material is linearly elastic, the constitutive equation is given by
\[
\bmsigma = \lambda\,\mathrm{tr}(\bmvarepsilon-\bmvarepsilon^g) {\bf I} + 2\mu(\bmvarepsilon-\bmvarepsilon^g)=\lambda\,\mathrm{tr}(\bmvarepsilon) {\bf I} + 2\mu(\bmvarepsilon)-\bmsigma^g,
\]
where the growth-induced stress tensor is defined as
\[
\bmsigma^g = \lambda\,\mathrm{tr}(\bmvarepsilon^g) {\bf I} + 2\mu\bmvarepsilon^g.
\]
Introducing the constant
\[
A := \lambda + 2\mu,
\]
the stress components can be written as
\[
\sigma_{rr} = Au'+\lambda\frac{u}{r}-\sigma^g_{rr}, \qquad
\sigma_{\theta\theta} = \lambda u'+A\frac{u}{r}-\sigma^g_{\theta\theta},
\qquad \sigma_{r\theta} = 0.
\]
If a  body force of magnitude \(p(r)\) acts in the radial direction, the equilibrium equation takes the form
\[
\frac{d\sigma_{rr}}{dr}+\frac{\sigma_{rr}-\sigma_{\theta\theta}}{r}+p=0.
\]
Moreover, assuming that the boundary \(\partial\Omega\) is traction-free, the boundary conditions are
\[
\sigma_{rr}(a)=0,\qquad \sigma_{rr}(b)=0.
\]
As in the previous section, we postulate that the growth strain tensor is determined by an optimality principle. More precisely, among all admissible growth configurations, the body selects the one that minimizes the work performed by the applied loads:
\[
W(u,\bmvarepsilon^g):=\int_a^b 2\pi r \,p(r)\,u(r)\,dr
\]
subject to the constraint
\[
\int_a^b r\bigl(\varepsilon^g_{rr}(r)+\varepsilon^g_{\theta\theta}(r)\bigr)\,dr=\Gamma.
\]
The growth problem  is:
\begin{equation}\label{pbm2d}
\begin{cases}
\displaystyle \inf_{u, \varepsilon^g}  & \displaystyle W(u,\bmvarepsilon^g):=\int_a^b 2\pi p(r)\,r\,u(r)\,dr,\\
&\mbox{subject to}\\
& \displaystyle \frac{d\sigma_{rr}}{dr}+\frac{\sigma_{rr}-\sigma_{\theta\theta}}{r}+p=0,\\
& \sigma_{rr}(a)=0\mbox{ and } \sigma_{rr}(b)=0,\\
&\bmsigma = \lambda\,\mathrm{tr}(\bmvarepsilon-\bmvarepsilon^g) {\bf I} + 2\mu(\bmvarepsilon-\bmvarepsilon^g),\\
& \displaystyle \varepsilon_{rr}=u', \qquad \varepsilon_{\theta\theta}=\frac{u}{r}, \qquad \varepsilon_{r\theta} = 0,\\
& \varepsilon^g_{rr}\ge 0\mbox{ and } \varepsilon^g_{\theta\theta}\ge 0,\\
& \displaystyle \int_a^b r\bigl(\varepsilon^g_{rr}(r)+\varepsilon^g_{\theta\theta}(r)\bigr)\,dr=\Gamma.
\end{cases}
\end{equation}

By using \eqref{pbm2d}$_4$ and \eqref{pbm2d}$_5$,
the equilibrium  equation \eqref{pbm2d}$_2$ reduces to
\[
u''+\frac{1}{r}u'-\frac{u}{r^2}=f_g(r)-\frac{p(r)}{A}, 
\]
where
\be
\label{eq-f}
f_g(r):=\frac{1}{A}\left[ \frac{d\sigma_{rr}^g}{dr}+\frac{\sigma_{rr}^g-\sigma_{\theta\theta}^g}{r}\right]=   \frac{1}{A}\left[A\, \frac{\partial\varepsilon^g_{rr}}{\partial r}+\lambda\,\frac{\partial\varepsilon^g_{\theta\theta}}{\partial r}
+\frac{2\mu}{r}\bigl(\varepsilon^g_{rr}-\varepsilon^g_{\theta\theta}\bigr)\right].
\ee
Taking into account the traction-free boundary conditions \eqref{pbm2d}$_3$, the displacement $u$ solves the problem:
$$
\begin{cases}
\displaystyle u''+\frac{1}{r}u'-\frac{u}{r^2}=f_g(r)-\frac{p(r)}{A} & \mbox{in }(a,b),\\
\displaystyle Au'+\lambda \frac u r =\sigma_{rr}^g & \mbox{on }\{a,b\}.
\end{cases}
$$
Let $u_p$ be the solution of
\be\label{ue}
\begin{cases}
\displaystyle u_p''+\frac{1}{r}u_p'-\frac{u_p}{r^2}=-\frac{p(r)}{A} & \mbox{in }(a,b),\\
\displaystyle Au_p'+\lambda \frac{u_p} r =0 & \mbox{on }\{a,b\},
\end{cases}
\ee
and $u_f$ the solution of
\be\label{ug}
\begin{cases}
\displaystyle u_f''+\frac{1}{r}u_f'-\frac{u_f}{r^2}=f_g(r) & \mbox{in }(a,b),\\
\displaystyle Au_f'+\lambda \frac{u_f} r =\sigma_{rr}^g & \mbox{on }\{a,b\}.
\end{cases}
\ee
Then
$$
u=u_p+u_f.
$$

As shown in the appendix, one finds that the solution $u_p$ can be expressed as
\be
u_p(r)=C_1^p\,r+\frac{C_2^p}{r}+\frac{r}{2}F_1^p(r)-\frac{1}{2r}F_2^p(r), 
\label{eq:up}
\ee
where
\[
F_1^p(r):=-\frac 1A\int_a^r p(s)\,ds,
\qquad
F_2^p(r):=-\frac 1A\int_a^r s^2 p(s)\,ds,
\]
\[
C_1^p=\frac{b^2S_b^p}{2(\mu+\lambda)(b^2-a^2)},
\qquad
C_2^p=\frac{a^2b^2 S_b^p}{2\mu(b^2-a^2)}, \qquad
S_b^p:=
-(\mu+\lambda)F_1^p(b)-\frac{\mu}{b^2}F_2^p(b).
\]

\subsection{Making the work done by the body-force explicit}

The work done by the body-force is
\[
W(u,\bmvarepsilon^g)=\int_a^b 2\pi p(r)\,r\,u(r)\,dr= \int_a^b 2\pi p(r)\,r\,u_p(r)\,dr+\int_a^b 2\pi p(r)\,r\,u_f(r)\,dr.
\]
Since $u_p$ is entirely determined by the body force $p$, the minimization of $W$ is equivalent, up to an additive constant, to minimizing the second integral. We therefore define
$$
W_g(u,\bmvarepsilon^g):=\int_a^b 2\pi p(r)\,r\,u_f(r)\,dr.
$$

It is shown in the appendix that $W_g$ can be written explicitly as
\be
W_g(u,\bmvarepsilon^g) =2\pi\int_a^b r \big[ \Sigma^p_{rr}(r) \veps_{rr}^g(r) + \Sigma^p_{\theta\theta}(r) \veps_{\theta\theta}^g(r)\big] dr,
\label{eq1}
\ee
where
$$
\Sigma^p_{rr}(r):=Au_p' +\lambda \frac{u_p}r, \qquad  \Sigma^p_{\theta\theta}(r):=\lambda u_p' +A\frac{u_p}{r}
$$
are the stress components associated with $u_p$, that is in the absence of growth.  
Note from \eqref{ue}$_2$ that
\be
\Sigma^p_{rr}(a)=\Sigma^p_{rr}(b)=0.
\label{eqab}
\ee
Finally, with the expression of $u_p$ from \eqref{eq:up}, we find
\be
\begin{aligned}
\Sigma^p_{rr}(r)&=(\mu+\lambda)\big(2C_1^p+F_1^p(r))+\mu  \frac{F_2^p(r)-2C_2^p}{r^2}\\
&=-\frac{(r^2-a^2)}{r^2(b^2-a^2)}\big((\mu+\lambda)b^2 F_1^p(b)+\mu F_2^p(b)\big)+(\mu+\lambda)F_1^p(r)+\frac{\mu}{r^2}F_2^p(r),\\
\Sigma^p_{\theta\theta}(r)&=(\mu+\lambda)\big(2C_1^p+F_1^p(r))-\mu  \frac{F_2^p(r)-2C_2^p}{r^2}\\
&=-\frac{(a^2+r^2)}{r^2(b^2-a^2)}\big((\mu+\lambda)b^2 F_1^p(b)+\mu F_2^p(b)\big)+(\mu+\lambda)F_1^p(r)-\frac{\mu}{r^2}F_2^p(r).
\end{aligned}
\label{Phigen}
\ee



%
%
%

\subsection{Growth under uniform load}

For $p(r)=p$, from \eqref{Phigen} we find
\be
\Sigma^p_{rr}(r)=p
\frac{(r-a)(b-r)(4\mu+3\lambda)\bigl(ab+(a+b)r\bigr)}{3r^2(a+b)(2\mu+\lambda)},
\label{eq11}
\ee
\be
\Sigma^p_{\theta\theta}(r)=p
\frac{(4 \mu + 3 \lambda) a^2b^2
+  (4\mu + 3 \lambda)(a^2 + ab + b^2) r^2
- (2\mu + 3 \lambda)(a+b)r^3}
{3r^2(a+b)(2\mu+\lambda)}. 
\label{eq111}
\ee

It can be verified that the functions  \(\Sigma^p_{rr}(r)\) and \(\Sigma^p_{\theta\theta}(r)\) have the following properties:
\be
\label{eq2}
\Sigma^p_{rr}(a) = \Sigma^p_{rr}(b) = 0, \qquad \frac{\Sigma^p_{rr}(r)}p > 0 \quad {\rm for} \quad a < r < b, 
\ee
\be
\label{eq3}
\frac{\Sigma^p_{\theta\theta}(r)-\Sigma^p_{rr}(r)}p
=
\frac{2\big[(4\mu+3\lambda)a^2b^2+\mu(a+b)r^3\big]}{3r^2(a+b)(2\mu+\lambda)} > 0  \quad {\rm for} \quad a \le r \le b, 
\ee
\be
\label{eq4}
\frac 1p \frac{d\Sigma^p_{\theta\theta}}{dr}
=
-\frac{2(4\mu+3\lambda)a^2b^2+(2\mu+3\lambda)(a+b)r^3}
{3r^3(a+b)(2\mu+\lambda)} <0 \quad {\rm for} \quad a \le r \le b, \quad \frac{\Sigma^p_{\theta\theta}(a)}p > \frac{\Sigma^p_{\theta\theta}(b)}p \ge 0. 
\ee

\subsubsection{Positive body-force: $\boldsymbol{p>0}$}

From \eqref{eq1}, using \eqref{pbm2d}$_6$, \eqref{eq2}, and  \eqref{eq3}, we deduce
\be\label{axyplow}
\begin{aligned}
W_g(u,\bmvarepsilon^g) & =2\pi\int_a^b r \big[ \Sigma^p_{rr}(r) \veps_{rr}^g(r) + \Sigma^p_{\theta\theta}(r) \veps_{\theta\theta}^g(r)\big] dr,\\
&\ge  2 \pi  \int_a^b r \big[ \Sigma^p_{rr}(r) \veps_{rr}^g(r) + \Sigma^p_{rr}(r) \veps_{\theta\theta}^g(r)\big] dr = 2 \pi  \int_a^b r \Sigma^p_{rr}(r)  \big[ \veps_{rr}^g(r) + \veps_{\theta\theta}^g(r)\big] dr\\
&\ge 2 \pi  \Sigma^p_{rr}(a) \int_a^b r   \big[ \veps_{rr}^g(r) + \veps_{\theta\theta}^g(r)\big] dr=2 \pi  \Sigma^p_{rr}(b) \int_a^b r   \big[ \veps_{rr}^g(r) + \veps_{\theta\theta}^g(r)\big] dr=0,
\end{aligned}
\ee
hence the lower bound is simply zero, that is,  if the bound is attained growth occurs without the loads performing work.
To see that the lower bound is attained, for any  $\eta\in (0,1)$ and for every integer $k$ let
$$\veps_{rr,k}^g=\eta \frac{k\Gamma t}{a+\frac 1{2k}} \chi_{[a,a+1/k]}+(1-\eta)\frac{k\Gamma t}{b-\frac 1{2k}} \chi_{[b-1/k,b]}\qquad \veps_{\theta\theta,k}^g=0.
$$ 
We note that
$$
\int_a^b r\bigl(\varepsilon^g_{rr,k}(r)+\varepsilon^g_{\theta\theta,k}(r)\bigr)\,dr=
\eta \frac{k\Gamma t}{a+\frac 1{2k}} \int_a^{a+1/k}r\,dr+(1-\eta)  \frac{k\Gamma t}{b-\frac 1{2k}} \int_{b-1/k}^br\,dr=\Gamma,
$$
hence \eqref{pbm2d}$_7$ is satisfied, and 
$$
\begin{aligned}
\lim_{k\to \infty}W_g(u,\bmvarepsilon^g_k) &=\lim_{k\to \infty}2\pi\Big(\eta \frac{k\Gamma t}{a+\frac 1{2k}} \int_a^{a+1/k}r\Sigma^p_{rr}(r)\,dr+(1-\eta)  \frac{k\Gamma t}{b-\frac 1{2k}} \int_{b-1/k}^br\Sigma^p_{rr}(r)\,dr\Big)\\
&=2\pi\big(\eta \Gamma t \Sigma^p_{rr}(a)+(1-\eta)  \Gamma t\Sigma^p_{rr}(b)\big)=0
\end{aligned}
$$
which shows that the lower bound \eqref{axyplow} is attained in the limit. Arguing as in Section \ref{meassol},  
we can check that the ``limit'' growth strains are measure-valued growth strains:

\begin{minipage}{0.2\textwidth}
 \hfill
 \begin{tikzpicture}[scale=0.5,>=Latex]

    \def\Ri{0.5}
    \def\Re{2.2}
     \def\Rii{0.3}
     \def\Ree{2.4}

    \fill[gray!15, even odd rule] (0,0) circle (\Re) (0,0) circle (\Ri);
     \fill[red, even odd rule] (0,0) circle (\Ri) (0,0) circle (\Rii);
     \fill[red, even odd rule] (0,0) circle (\Ree) (0,0) circle (\Re);
    \draw[thick] (0,0) circle (\Re);
    \draw[thick] (0,0) circle (\Ri);

    \foreach \r in {0.8,1.6} {
        \foreach \a in {0,30,...,330} {
            \draw[->,thick]
                ({(\r-0.15)*cos(\a)},{(\r-0.15)*sin(\a)})
                --
                ({(\r+0.55)*cos(\a)},{(\r+0.55)*sin(\a)});
        }
    }

\end{tikzpicture}
 \end{minipage}
 \qquad\qquad
  \begin{minipage}{0.8\textwidth}
\begin{flushleft}
\(
\begin{aligned}
\veps_{\theta\theta}^g &= 0, \\
\veps_{rr}^g &= \eta \frac{\Gamma}{a} \delta_a +(1-\eta) \frac{\Gamma}{b} \delta_b  \quad \forall \eta\in (0,1), 
\end{aligned}
\)
\end{flushleft}
 \end{minipage}

\noindent
that is, the growth strain in the radial direction is concentrated at the inner and outer boundaries.
We further note that the solution is not unique. Indeed, growth localized at \( r=b \) effectively extends the annulus outward, while growth localized at \( r=a \) partially fills the internal cavity. In both cases, the added material does not induce displacement in the body (in addition to that produced by the load $p$), and therefore does not contribute to the mechanical work. As a consequence, different distributions of growth concentrated at the boundaries may yield the same value of the objective functional.

\subsubsection{Negative body-force: $\boldsymbol{p<0}$}

From \eqref{eq1}, using \eqref{pbm2d}$_6$, \eqref{eq3}, and  \eqref{eq4}, we deduce
\be
\begin{aligned}
W_g(u,\bmvarepsilon^g) & =2\pi\int_a^b r \big[ \Sigma^p_{rr}(r) \veps_{rr}^g(r) + \Sigma^p_{\theta\theta}(r) \veps_{\theta\theta}^g(r)\big] dr,\\
 &\ge 2 \pi  \int_a^b r \big[ \Sigma^p_{\theta\theta}(r) \veps_{rr}^g(r) + \Sigma^p_{\theta\theta}(r) \veps_{\theta\theta}^g(r)\big] dr = 2 \pi  \int_a^b r \Sigma^p_{\theta\theta}(r)  \big[ \veps_{rr}^g(r) + \veps_{\theta\theta}^g(r)\big] dr \\
&\ge
2 \pi  \Sigma^p_{\theta\theta}(a) \int_a^b r  \big[ \veps_{rr}^g(r) + \veps_{\theta\theta}^g(r)\big] dr.
\end{aligned} 
\ee
Hence, by \eqref{pbm2d}$_7$, we have that
$$
W_g(u,\bmvarepsilon^g) \ge 2 \pi  \, \Gamma  t \, \Sigma^p_{\theta\theta}(a).
$$
By taking 
$$\veps_{rr,k}^g=0\qquad \veps_{\theta\theta,k}^g=\frac{k\Gamma t}{a+\frac 1{2k}} \chi_{[a,a+1/k]}
$$ 
and
arguing as in the previous section,  we can check that the lower bound is attained in the limit,
and the ``limit'' growth strains are:

 \begin{minipage}{0.4\textwidth}
 \hfill
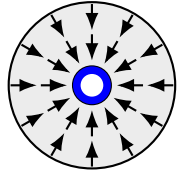
 \begin{tikzpicture}[scale=0.5,>=Latex]

    \def\Ri{0.5}
    \def\Re{2.2}
    \def\Rii{0.3}

    \fill[gray!15, even odd rule] (0,0) circle (\Re) (0,0) circle (\Ri);
    \draw[thick] (0,0) circle (\Re);
    \draw[thick] (0,0) circle (\Ri);
     \fill[blue, even odd rule] (0,0) circle (\Ri) (0,0) circle (\Rii);

    \foreach \r in {0.8,1.6} {
        \foreach \a in {0,30,...,330} {
            \draw[<-,thick]
                ({(\r-0.15)*cos(\a)},{(\r-0.15)*sin(\a)})
                --
                ({(\r+0.55)*cos(\a)},{(\r+0.55)*sin(\a)});
        }
    }

\end{tikzpicture}
 \end{minipage}
 \qquad\qquad
  \begin{minipage}{0.6\textwidth}
\begin{flushleft}
\(
\begin{aligned}
\veps_{rr}^g &= 0,\\
\veps_{\theta\theta}^g&= \frac{\Gamma}{a} \delta_a;
\end{aligned}
\)
\end{flushleft}
 \end{minipage}

 \noindent
that is, the growth strain in the hoop direction is concentrated at the inner boundary.
Indeed, growth in the radial direction localized at the boundary does not contribute to the work of the applied loads, as observed in the previous case. Conversely, growth in the hoop direction generates displacements that result in negative work. Consequently, the optimal strategy is to localize the hoop growth strain at the inner boundary.

\subsection{Growth under a sign-changing load}

%
%
%
%

Let
$$
c=\frac{a+b}2
$$
and consider the load
$$
p(r)=\begin{cases}
+p & \mbox{if }r\in(a,c)\\
-p & \mbox{if }r\in(c,b).
\end{cases}
$$
Then,  from \eqref{Phigen} we find
$$
\Sigma^p_{rr}(r)=\begin{cases}
\displaystyle p\big(-k_{r1} r+ k_{2} + k_{3}/r^2\big)  & \mbox{if }r\in(a,c),\\
\displaystyle p\big(+k_{r1} r- k_{4} - k_{5}/r^2\big)  & \mbox{if }r\in(c,b),
\end{cases}
$$
and 
$$
\Sigma^p_{\theta\theta}(r)=\begin{cases}
\displaystyle p\big(-k_{\theta1} r+ k_{2} - k_{3}/r^2\big)  & \mbox{if }r\in(a,c),\\
\displaystyle p\big(+k_{\theta1} r- k_{4} + k_{5}/r^2\big)  & \mbox{if }r\in(c,b),
\end{cases}
$$
where
$$
k_{r1}:=\frac{4\mu+3\lambda}{3(2\mu+\lambda)}, \qquad k_{2}:=\frac{(5a-b)\mu+4a\lambda}{4(2\mu+\lambda)},\qquad k_{3}:=\frac{a^2(a+3b)\mu}{12(2\mu+\lambda)},
$$
$$
k_{\theta1}:=\frac{2\mu+3\lambda}{3(2\mu+\lambda)}, \qquad k_{4}:=\frac{ (5b-a)\mu+4b\lambda}{4(2\mu+\lambda)}, \qquad k_{5}:=\frac{b^2(3a+b)\mu}{12(2\mu+\lambda)}.
$$
We note that the constants $k_{r1}$, $k_{\theta1}$, $k_{3}$, $k_{4}$, and $k_{5}$ are positive. It follows that $\Sigma^p_{rr}/p$ is strictly decreasing on $(a,c)$, being the sum of a constant term and two strictly decreasing functions. Similarly, $\Sigma^p_{rr}/p$ is strictly increasing on $(c,b)$. Hence
$$
\min_{r\in (a,b)} \frac{\Sigma^p_{rr}(r)}p=\frac{\Sigma^p_{rr}(c)}p=-(b-a)\frac{6(a+b)^2\lambda+(11a^2+26 ab +11b^2)\mu}{12 (a+b)^2 (2\mu+\lambda)},
$$
and
$$
\max_{r\in(a,b)} \frac{\Sigma^p_{rr}(r)}p=\frac{\Sigma^p_{rr}(a)}p=\frac{\Sigma^p_{rr}(b)}p=0,
$$
where we used \eqref{eqab}.

Setting
$$
L_\theta(r):= \begin{cases}
\displaystyle -k_{\theta1} c+ k_{2} - k_{3}/a^2  & \mbox{in }(a,c),\\
\displaystyle +k_{\theta1} c- k_{4} + k_{5}/b^2  & \mbox{in }(c,b),
\end{cases}
\quad\mbox{and}\quad
U_\theta(r):= \begin{cases}
\displaystyle -k_{\theta1} a+ k_{2} - k_{3}/c^2  & \mbox{in }(a,c),\\
\displaystyle +k_{\theta1} b- k_{4} + k_{5}/c^2  & \mbox{in }(c,b),
\end{cases}
$$
we have that
$$
L_\theta(r)\le \frac{\Sigma^p_{\theta\theta}(r)}p \le U_\theta(r)\qquad \forall r\in (a,b).
$$
It turns out that
$$
L_\theta(r)=-\frac{(b-a)(5\mu+3\lambda)}{6(2\mu+\lambda)}
\quad\mbox{and}\quad
U_\theta(r)=-\frac{(b-a)(3a^2+2ab+3b^2)\mu}{12(a+b)^2(2\mu+\lambda)}
\qquad \forall r\in (a,b),
$$
and
$$
\frac{\Sigma^p_{rr}(c)}p-L_\theta(r)=-\mu\frac{b^3-a^3+5ab(b-a)}{12(a+b)^2(2\mu+\lambda)}\qquad \forall r\in (a,b).
$$
Therefore
\begin{equation}\label{bounds}
\min_{r\in (a,b)} \frac{\Sigma^p_{rr}(r)}p=\frac{\Sigma^p_{rr}(c)}p< L_\theta(r)\le \frac{\Sigma^p_{\theta\theta}(r)}p \le U_\theta(r)< 0 \le \max_{r\in(a,b)} \frac{\Sigma^p_{rr}(r)}p,\quad \forall r\in (a,b).
\end{equation}

\subsubsection{The load $\boldsymbol{p(r)}$ positive in $\boldsymbol{(a,c)}$ and negative in $\boldsymbol{(c,b)}$}

Let $p>0$. From \eqref{bounds} we deduce that  $\min_{r\in (a,b)} \Sigma^p_{\theta\theta}>\min_{r\in (a,b)} \Sigma^p_{rr}$, and therefore
$$
\begin{aligned}
W_g(u,\bmvarepsilon^g) &\ge 2\pi\int_a^b r \big[ \min_{r\in (a,b)} \Sigma^p_{rr}\, \veps_{rr}^g(r) + \min_{r\in (a,b)} \Sigma^p_{\theta\theta}\,\veps_{\theta\theta}^g(r)\big] dr \\
&\ge 2\pi \min_{r\in (a,b)} \Sigma^p_{rr} \int_a^b r \big[ \veps_{rr}^g(r) + \veps_{\theta\theta}^g(r)\big] dr\\
&=2\pi \Sigma^p_{rr}(c) \Gamma.
\end{aligned}
$$
Also in this case it is possible to check that
the lower bound is attained through a limit process,
and the ``limit'' growth strains are:

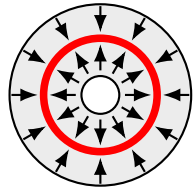
\begin{minipage}{0.4\textwidth}
 \hfill
 \begin{tikzpicture}[scale=0.5,>=Latex]

        \def\Ri{0.5}
    \def\Re{2.2}
     \def\Rii{0.3}
     \def\Ree{2.4}
          \def\Rm{1.4}
     \def\Rmm{1.6}

    \fill[gray!15, even odd rule] (0,0) circle (\Ree) (0,0) circle (\Ri);
         \fill[red, even odd rule] (0,0) circle (\Rmm) (0,0) circle (\Rm);

    \draw[thick] (0,0) circle (\Ree);
    \draw[thick] (0,0) circle (\Ri);

    \foreach \r in {0.8} {
        \foreach \a in {0,30,...,330} {
            \draw[->,thick]
                ({(\r-0.15)*cos(\a)},{(\r-0.15)*sin(\a)})
                --
                ({(\r+0.55)*cos(\a)},{(\r+0.55)*sin(\a)});
        }
    }
\foreach \r in {1.8} {
        \foreach \a in {0,30,...,330} {
            \draw[<-,thick]
                ({(\r-0.15)*cos(\a)},{(\r-0.15)*sin(\a)})
                --
                ({(\r+0.55)*cos(\a)},{(\r+0.55)*sin(\a)});
        }
    }

\end{tikzpicture}
 \end{minipage}
 \qquad\qquad
  \begin{minipage}{0.8\textwidth}
\begin{flushleft}
\(
\begin{aligned}
\veps_{\theta\theta}^g &= 0, \\
\veps_{rr}^g &= \frac{\Gamma}{c} \delta_c.
\end{aligned}
\)
\end{flushleft}
 \end{minipage}

Growth concentrated at the mid-radius of the annulus generates an internal expansion that drives the two portions of the annulus toward the boundaries. As a result, the induced deformation is converted as efficiently as possible into displacements opposing the external loads. Consequently, the loads perform the maximum negative work.

\subsubsection{The load $\boldsymbol{p(r)}$ negative in $\boldsymbol{(a,c)}$ and positive in $\boldsymbol{(c,b)}$}

Let $p<0$. From \eqref{bounds} we deduce that 
$$
\Sigma^p_{\theta\theta}(r)> p\max_{r\in(a,b)} \frac{\Sigma^p_{rr}(r)}p=\min_{r\in(a,b)} \Sigma^p_{rr}(r),\quad \forall r\in (a,b),
$$
hence $\min_{r\in (a,b)} \Sigma^p_{\theta\theta}>\min_{r\in (a,b)} \Sigma^p_{rr}$ and
$$
\begin{aligned}
W_g(u,\bmvarepsilon^g) &\ge 2\pi\int_a^b r \big[ \min_{r\in (a,b)} \Sigma^p_{rr}\, \veps_{rr}^g(r) + \min_{r\in (a,b)} \Sigma^p_{\theta\theta}\,\veps_{\theta\theta}^g(r)\big] dr \\
&\ge 2\pi \min_{r\in (a,b)} \Sigma^p_{rr} \int_a^b r \big[ \veps_{rr}^g(r) + \veps_{\theta\theta}^g(r)\big] dr\\
&=0.
\end{aligned}
$$
Again, it is possible to check that
the lower bound is attained through a limit process,
and the ``limit'' growth strains are:

\begin{minipage}{0.2\textwidth}
 \hfill
 \begin{tikzpicture}[scale=0.5,>=Latex]

    \def\Ri{0.5}
    \def\Re{2.2}
     \def\Rii{0.3}
     \def\Ree{2.4}

    \fill[gray!15, even odd rule] (0,0) circle (\Re) (0,0) circle (\Ri);
     \fill[red, even odd rule] (0,0) circle (\Ri) (0,0) circle (\Rii);
     \fill[red, even odd rule] (0,0) circle (\Ree) (0,0) circle (\Re);
    \draw[thick] (0,0) circle (\Re);
    \draw[thick] (0,0) circle (\Ri);

\draw[dashed] (0,0) circle [radius=1.5];

    \foreach \r in {0.8} {
        \foreach \a in {0,30,...,330} {
            \draw[<-,thick]
                ({(\r-0.15)*cos(\a)},{(\r-0.15)*sin(\a)})
                --
                ({(\r+0.55)*cos(\a)},{(\r+0.55)*sin(\a)});
        }
    }
\foreach \r in {1.8} {
        \foreach \a in {0,30,...,330} {
            \draw[->,thick]
                ({(\r-0.15)*cos(\a)},{(\r-0.15)*sin(\a)})
                --
                ({(\r+0.55)*cos(\a)},{(\r+0.55)*sin(\a)});
        }
    }

\end{tikzpicture}
 \end{minipage}
 \qquad\qquad
  \begin{minipage}{0.8\textwidth}
\begin{flushleft}
\(
\begin{aligned}
\veps_{\theta\theta}^g(r) &= 0, \\
\veps_{rr}^g(r) &= \eta \frac{\Gamma}{a} \delta_a +(1-\eta) \frac{\Gamma}{b} \delta_b  \quad \forall \eta\in (0,1).
\end{aligned}
\)
\end{flushleft}
 \end{minipage}


\section{Conclusions}

The central message of this work is that volumetric growth, when driven by the minimization of the work performed by external loads, may naturally select growth configurations that are effectively of surface type. Although growth is introduced as a distributed volumetric mechanism through admissible growth strain fields, the optimal solutions are not distributed. Instead, they concentrate as measures supported on lower-dimensional subsets of the body.

For the one-dimensional bar, the optimal growth strain collapses to a Dirac measure located at the point where the axial force attains its minimum value. Depending on the loading, this point may lie on the boundary or in the interior of the body. In the case of a uniform load, volumetric growth localizes at the boundary and becomes indistinguishable from a surface-growth process. 

The axisymmetric annular problem reveals the same underlying mechanism. Under uniform radial loads, the optimal growth strains concentrate at the boundaries of the annulus. Positive body forces lead to radial growth localized at the inner and outer surfaces, while negative body forces select hoop growth concentrated at the inner boundary. When the load changes sign across the thickness, the optimal growth localizes at an intermediate radius, producing an internal interface that separates regions displacing in opposite directions. In every case, the optimal growth is singular and supported on lower-dimensional subsets of the body.

Our analysis shows that the direction and location of concentration is determined as follows: if $\min_{r\in [a,b]} \Sigma_{rr}^p(r)<\min_{r\in [a,b]} \Sigma_{\theta\theta}^p(r)$ concentration occurs in the radial direction  ($\epsg_{\theta\theta}=0$)  and it is located at any minimum point of $ \Sigma_{rr}^p$. While, if $\min_{r\in [a,b]} \Sigma_{rr}^p(r)>\min_{r\in [a,b]} \Sigma_{\theta\theta}^p(r)$ growth occurs in the circumferential direction. 
Growth is essentially controlled by
$$
\min\big\{\min_{r\in [a,b]} \Sigma_{rr}^p(r),\min_{r\in [a,b]} \Sigma_{\theta\theta}^p(r)\big\}.
$$

These results show that surface growth can emerge as the  limit of a volumetric growth process, a possibility that was already anticipated in the pioneering work of Di Carlo \cite{DiCarlo2005}. The dimensional reduction is not imposed a priori; rather, it is selected by the optimality criterion itself. From this perspective, surface growth may be interpreted as the asymptotic manifestation of volumetric growth when the system seeks the most efficient way to modify its configuration under mechanical loading.

More broadly, the analysis highlights a variational mechanism capable of explaining why growth  is often observed to occur preferentially at boundaries, interfaces, or highly localized regions. The emergence of measure-valued growth distributions indicates that the interplay between mechanics and optimality can drive volumetric growth toward singular structures, thereby providing a mathematical bridge between distributed growth models and classical surface-growth descriptions. In this sense, the present results offer a variational explanation for the emergence of surface growth from bulk growth processes and further clarifies the mechanical conditions under which volumetric growth selects surface growth.


\section{Appendix}\label{sec:appendix}

Here we show the details of the calculations that led to the results we stated previously in \eqref{eq:up} and \eqref{eq1}.

\subsection{Determination of the displacement field $\boldsymbol{u_p}$:}

The general solution $u_p$ of \eqref{ue} can be expressed as
\[
u_p(r)=C_1^p\,r+\frac{C_2^p}{r}+\frac{r}{2}F_1^p(r)-\frac{1}{2r}F_2^p(r), 
\]
where
\[
F_1^p(r):=-\frac 1A\int_a^r p(s)\,ds,
\qquad
F_2^p(r):=-\frac 1A\int_a^r s^2 p(s)\,ds.
\]
The traction-free boundary conditions therefore become
\begin{equation}\label{tfup}
\begin{cases}
\displaystyle 2(\mu+\lambda)C_1^p-\frac{2\mu}{a^2}C_2^p=0,\\[8pt]
\displaystyle 2(\mu+\lambda)C_1^p-\frac{2\mu}{b^2}C_2^p=S_b^p, 
\end{cases}
\end{equation}
where
$$
S_b^p:=
-(\mu+\lambda)F_1^p(b)-\frac{\mu}{b^2}F_2^p(b).
$$
From \eqref{tfup} we find
\[
C_1^p=\frac{b^2S_b^p}{2(\mu+\lambda)(b^2-a^2)},
\qquad
C_2^p=\frac{a^2b^2 S_b^p}{2\mu(b^2-a^2)}.
\]


\subsection{Determination of the displacement field $\boldsymbol{u_f}$:}

The solution $u_f$ of \eqref{ug} can be expressed as
\[
u_f(r)=C_1^f\,r+\frac{C_2^f}{r}+\frac{r}{2}F_1^f(r)-\frac{1}{2r}F_2^f(r),
\]
where
\[
F_1^f(r):=\int_a^r f_g(s)\,ds,
\qquad
F_2^f(r):=\int_a^r s^2 f_g(s)\,ds.
\]
The  boundary conditions can be written as
\begin{equation}\label{tfuf}
\begin{cases}
\displaystyle 2(\mu+\lambda)C_1^f-\frac{2\mu}{a^2}C_2^f=S_a^f,\\[8pt]
\displaystyle 2(\mu+\lambda)C_1^f-\frac{2\mu}{b^2}C_2^f=S_b^f, 
\end{cases}
\end{equation}
where
\be
S_a^f:=(2\mu+\lambda)\varepsilon^g_{rr}(a)+\lambda\varepsilon^g_{\theta\theta}(a), 
\label{eq-Sa}
\ee
\be
S_b^f:=(2\mu+\lambda)\varepsilon^g_{rr}(b)+\lambda\varepsilon^g_{\theta\theta}(b)
-(\mu+\lambda)I_1^f-\frac{\mu}{b^2}I_2^f, 
\label{eq-Sb}
\ee
with
\be
\label{eq-I1I2}
I_1^f := F_1^f(b)=\int_a^b f_g(r)\,dr,
\qquad
I_2^f :=F_2^f(b) = \int_a^b r^2f_g(r)\,dr.
\ee
Solving \eqref{tfuf} leads to
\[
C_1^f=\frac{b^2S_b^f-a^2S_a^f}{2(\mu+\lambda)(b^2-a^2)},
\qquad
C_2^f=\frac{a^2b^2}{2\mu(b^2-a^2)}\bigl(S_b^f-S_a^f\bigr). 
\]
Therefore the displacement field $u_f$ is
\[
 u_f(r)=\frac{b^2S_b^f-a^2S_a^f}{2(\mu+\lambda)(b^2-a^2)}\,r
 +\frac{a^2b^2\bigl(S_b^f-S_a^f\bigr)}{2\mu(b^2-a^2)}\,\frac{1}{r}
 +\frac{r}{2}F_1^f(r)-\frac{1}{2r}F_2^f(r). 
\]


\subsection{Determination of the work $\boldsymbol{W_g}$:}

 Let $w$ and $v$ be two twice differentiable functions on $(a,b)$. Upon integrating by parts we find the following identity
$$
\int_a^b \big(v''+\frac{v'}r-\frac{v}{r^2}\big) wr\,dr=\big[v'wr\big]_a^b-\big[vw'r\big]_a^b+\int_a^b \big(w''+\frac{w'}r-\frac{w}{r^2}\big) vr\,dr.
$$
On taking $v = u_p,$ $w = u_f$ and using \eqref{ue}, \eqref{ug}, we find
$$
\int_a^b -\frac{p(r)}{A} u_f r\,dr=\big[u_p' u_f r\big]_a^b-\big[u_p u_f'r\big]_a^b+\int_a^b f_g u_p r\,dr,
$$
and
$$
W_g(u,\bmvarepsilon^g)= -2\pi A \Big(\big[u_p' u_f r\big]_a^b-\big[u_p u_f'r\big]_a^b+\int_a^b f_g u_p r\,dr\Big).
$$
Still from \eqref{ue} and \eqref{ug} we deduce that
$$
u_p'=-\frac{\lambda}{A} \frac{u_p} r    \qquad  u_f'=-\frac{\lambda}{A} \frac{u_f} r+\frac{1}{A}\sigma_{rr}^g \qquad \mbox{on }\{a,b\},
$$
hence
$$
\big[u_p' u_f r\big]_a^b-\big[u_p u_f'r\big]_a^b=-\frac{\lambda}{A}\big[u_p u_f \big]_a^b+\frac{\lambda}{A}\big[u_p u_f\big]_a^b-\frac{1}{A}\big[u_p \sigma_{rr}^g r\big]_a^b=-\frac{1}{A}\big[u_p \sigma_{rr}^g r\big]_a^b,
$$
and therefore
\be
W_g(u,\bmvarepsilon^g)= -2\pi  \Big(-\big[u_p \sigma_{rr}^gr\big]_a^b+A\int_a^b f_g u_p r\,dr\Big).
\label{eq001}
\ee
The integral appearing in \eqref{eq001}, can be rewritten, using   the definition of $f_g$, as
$$
\begin{aligned}
A\int_a^b f_g u_p r\,dr&= \int_a^b \left[ \frac{d\sigma_{rr}^g}{dr}+\frac{\sigma_{rr}^g-\sigma_{\theta\theta}^g}{r}\right] u_p r\,dr\\
&=\big[u_p \sigma_{rr}^gr\big]_a^b +\int_a^b  [ -\sigma_{rr}^g (u_p' r+u_p)+(\sigma_{rr}^g-\sigma_{\theta\theta}^g)u_p ] \,dr\\
&=\big[u_p \sigma_{rr}^gr\big]_a^b -\int_a^b  [\sigma_{rr}^g u_p' r+\sigma_{\theta\theta}^gu_p]  \,dr.
\end{aligned}
$$
Thus, from  \eqref{eq001} we find
$$
W_g(u,\bmvarepsilon^g)= 2\pi\int_a^b  [\sigma_{rr}^g u_p' r+\sigma_{\theta\theta}^gu_p ] \,dr,
$$
which on using the constitutive equation $\bmsigma^g = \lambda\,\mathrm{tr}(\bmvarepsilon^g) {\bf I} + 2\mu\bmvarepsilon^g$ yields \eqref{eq1}.

\medskip

\noindent
{\bf Acknowledgments.} We gratefully acknowledge the support of the MIT-UNIPI Seed Fund.  R.P.  and  M.P.S. thankfully acknowledge the support of the Italian National Group of Mathematical Physics (GNFM-INdAM).

\end{document}